\documentclass[aps,prl,twocolumn,groupedaddress]{revtex4-1}
\usepackage{graphicx, amsmath}
\usepackage{xcolor}
\usepackage{float}					
\graphicspath{ {Figures/}}
\begin{document}


\title{Vortex phases and glassy dynamics in the highly anisotropic superconductor HgBa$_2$CuO$_{4+\delta}$}

\author{Serena Eley}
\email[]{serenaeley@mines.edu}
\affiliation{Department of Physics, Colorado School of Mines, Golden, CO 80401}
\affiliation{Los Alamos National Laboratory, Los Alamos, NM 87545}

\author{Roland Willa}
\affiliation{Institute for Theory of Condensed Matter, Karlsruhe Institute of Technology, Karlsruhe, Germany}

\author{Mun K. Chan}
\affiliation{Pulsed Field Facility, National High Magnetic Field Laboratory, Los Alamos National Laboratory, Los Alamos, NM 87545}

\author{Eric D. Bauer}
\affiliation{Los Alamos National Laboratory, Los Alamos, NM 87545}

\author{Leonardo Civale}
\affiliation{Los Alamos National Laboratory, Los Alamos, NM 87545}

\date{\today}

\begin{abstract}
We present an extensive study of vortex dynamics in a high-quality single crystal of HgBa$_2$CuO$_{4+\delta}$ (Hg1201), a highly anisotropic superconductor that is a model system for studying the effects of anisotropy. From magnetization $M$ measurements over a wide range of temperatures $T$ and fields $H$, we construct a detailed vortex phase diagram. We find that the temperature-dependent vortex penetration field $H_p(T)$, second magnetization peak $H_{smp}(T)$, and irreversibility field $H_{irr}(T)$ all decay exponentially at low temperatures and exhibit an abrupt change in behavior at high temperatures $T/T_c \gtrsim 0.5$.  By measuring the rates of thermally activated vortex motion (creep) $S(T,H)=|d \ln M(T,H) / d \ln t|$, we reveal glassy behavior involving collective creep of bundles of 2D pancake vortices as well as temperature- and time-tuned crossovers from elastic (collective) dynamics to plastic flow.  Based on the creep results, we show that the second magnetization peak coincides with the  elastic-to-plastic crossover at low $T$, yet the mechanism changes at higher temperatures.

\vspace{0.3cm}


\end{abstract}


\maketitle

\section{I. Introduction}

Interest in copper-oxide superconductors (cuprates) is fueled by their technological potential and the outstanding mystery of the mechanism governing high-temperature superconductivity, which stifles prediction of new superconductors.  It is known that, in cuprates, superconductivity is hosted in the crystallographic ab-planes. This induces anisotropy $\gamma$ between the in-plane (ab) and out-of-plane (c-axis) fundamental superconducting parameters, such as the penetration depth $\lambda_{ab}=\lambda_c/\gamma$ and coherence length $\xi_{ab}=\gamma\xi_c$.  When evaluating the potential of superconductors for technological applications, high anisotropy compels considerations beyond the typical metrics of high critical temperature $T_c$, critical current density  $J_c$, and upper critical field $H_{c2}$.  This is because thermal fluctuations profoundly impact anisotropic materials' electronic and magnetic properties, which are significantly influenced by the dynamics of vortices.  Consequently, thermally activated vortex motion (creep) is fast and $J_c$ vanishes at an irreversibility field $H_{irr}$ that can be much less than $H_{c2}$, potentially negating the otherwise advantageous properties of these materials. Understanding vortex dynamics in cuprates is not only technologically relevant, but also can substantially contribute to the debate over the degree to which superconductivity in cuprates is conventional \cite{Berthod2017}.




Magnetic flux penetrates superconductors immersed in fields greater than the lower critical field $H_{c1}$.  This does not quench superconductivity in high-$T_c$ materials provided that the field remains below $H_{c2}$. In layered cuprates, interior flux can appear as stacks of weakly-coupled 2D pancake vortices, each localized on a Cu-O plane.  Pancake vortices within a stack are not necessarily aligned and interact both magnetically (owing to their moments) and through Josephson coupling between pancakes in adjacent planes.  If this coupling is sufficiently strong, the stacks may behave as continuous strings, hence be considered 3D vortex lines. The differing dynamics of 2D pancakes and 3D vortex lines should therefore play a major role in determining the phase diagram in highly anisotropic materials.


The superconductor HgBa$_2$CuO$_{4+\delta}$ (Hg1201) is recognized as ideal for systematically studying the effects of high anisotropy.  This is because its clean microstructure enables probing intrinsic, rather than sample-dependent, properties associated with high anisotropy \cite{Barisic2008}.  Specifically, Hg1201 crystals do not contain common defects, such as twin-boundaries and rare-earth-oxide precipitates \cite{Wagner1993,Pelloquin1997,Viallet1997}.  Furthermore, it has a simple tetragonal structure and optimally doped Hg1201 has the highest $T_c$ among single Cu-O layer materials, permitting thorough studies of the effects of thermal fluctuations on the superconducting state. Despite these desirable characteristics, the paucity of research on Hg1201 results from the challenges of growing large, high-quality single crystals. 





\begin{figure*}
\centering
\includegraphics[width=0.95\textwidth]{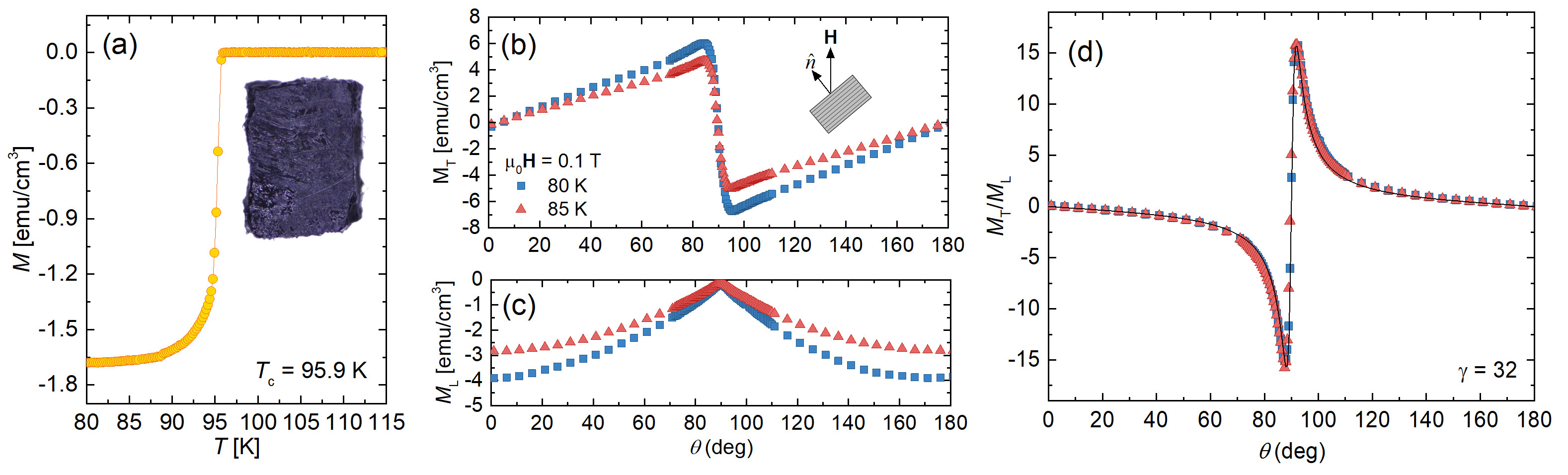}
\caption{(a) Temperature dependent magnetization $M(T)$ measured at $H =$5 Oe after zero-field cooling, revealing $T_c \approx 95.9 \textnormal{ K}$, consistent with expectations for optimally doped Hg1201.  Angular dependence of the (b) transverse ($M_T$) and (c) longitudinal ($M_L$) components of the magnetization, and (d) the ratio $M_T/M_L$ in an applied field of 0.1 T and temperatures 80 K and 85 K. Note that in (d), the data for the two temperatures overlap and the black curve is a fit of the 80 K data to Eq.\ \eqref{eq:Anisotropy} that yields an anisotropy factor $\gamma \approx 32$. \label{Figure1}}
\end{figure*}

In this paper, we construct a detailed phase diagram of vortex dynamics in a clean, optimally doped Hg1201 single crystal.  We find that the temperature-dependent vortex penetration field $H_p(T)$, second magnetization peak $H_{smp}(T)$, and irreversibility field $H_{irr}(T)$ all decay exponentially at low temperatures and exhibit an abrupt change in behavior at high temperatures.  We present complementary vortex creep measurements over a wide range of the phase diagram that reveal the broad extent to which the dynamics of pancake vortices determine the magnetic properties in our sample.  Our main findings from these measurements are as follows: First, the crystal hosts a vortex glass state characterized by collective creep of large bundles of pancake vortices at low temperatures $T/T_c \leq 0.4$ and applied fields $\mu_0H < 0.5 \textnormal{ T}$. The glass state persists at higher fields, yet the bundle size shrinks.  Second, we find temperature-tuned crossovers from elastic (collective) dynamics to plastic flow. By sitting near the crossover temperature and measuring over an extended time frame, we additionally capture a transition from elastic to plastic dynamics over time.  Last, we show that the second magnetization peak does not originate from elastic-to-plastic crossovers over most of the phase diagram; these crossovers  only coincides with the second magnetization peak at low temperatures $T/T_c<0.2$.

\section{II. Experimental Details}

Hg1201 single crystals were grown using an encapsulated self-flux method \cite{zhao06} at Los Alamos National Laboratory. The crystals were subsequently heat-treated at 350$ ^o$C in air and quenched to room temperature to achieve near optimal doping \cite{yamamoto00}. The high-quality of the synthesized crystals is evinced by the observation of large quantum oscillations in other samples from the same growth batch \cite{chan18}.  Multiple crystals were measured to verify reproducibility. The results presented in this manuscript were collected on a crystal with dimensions $1.28 \times 0.84 \times 0.24$ mm$^3$ and mass of 1.9 mg, shown in the Fig.\ \ref{Figure1}(a) inset. 

All measurements were performed using a Quantum Design superconductor quantum interference device (SQUID) magnetometer equipped with two independent sets of detection coils to measure the magnetic moment in the direction of ($m_L$) and transverse to ($m_T$) the applied magnetic field.  For measurements requiring manipulating the field orientation, the crystal was placed on a rotating sample mount.  Most measurements, however, were conducted with the field aligned with the sample c-axis ($ H \parallel c$), in which case the sample was mounted on a delrin disk inside a straw. 

\section{III. Results and Discussion}
\subsection{A. Critical Temperature and Anisotropy}
Temperature-dependent magnetization measurements $M(T)$ in a field of 5 Oe yielded a critical temperature $T_c \approx 95.9 \textnormal{ K}$ consistent with near optimal doping \cite{Barisic2008, yamamoto00} , see Fig.\ \ref{Figure1}(a). To determine the anisotropy, we measured the ratio of the transverse ($M_{T}$) to the longitudinal ($M_{L}$) magnetization at various field orientations ($\theta$)
relative to the c-axis in the reversible (vortex liquid) regime.  The raw data is plotted in Figs.\ \ref{Figure1}(b) and \ref{Figure1}(c). As shown in Fig.\ \ref{Figure1}(d), a least squares fit of the data to Eq.\ \eqref{eq:Anisotropy}, the Kogan model \cite{Kogan1988, Mosqueira2010}, 

\begin{equation} \label{eq:Anisotropy}
\frac{M_T}{M_L}=(1-\gamma^2)\frac{\sin \theta \cos \theta}{\sin^2 \theta + \gamma^2 \cos^2 \theta},
\end{equation}

\noindent produces an anisotropy of $\gamma \approx 32$.  This is consistent with previous work on optimally doped Hg1201 single crystals.  Specifically, angle dependent torque magnetometry studies \cite{Xia2012,Hofer2000,Hofer1998,Hofer1999} found $\gamma \approx 27-30$.  Additionally, a study \cite{Bras1996} that measured the magnetization at two field orientations, perpendicular ($M_\perp$) and parallel ($M_\parallel$) to the CuO$_2$ planes%
, found $\gamma \approx 30$ using a self-consistency equation from anisotropic Ginzburg-Landau theory $M_\perp (H)=\gamma M_\parallel (\gamma H)$.

\subsection{B. Irreversible Magnetization}

\begin{figure}
\centering
\includegraphics[width=0.45\textwidth]{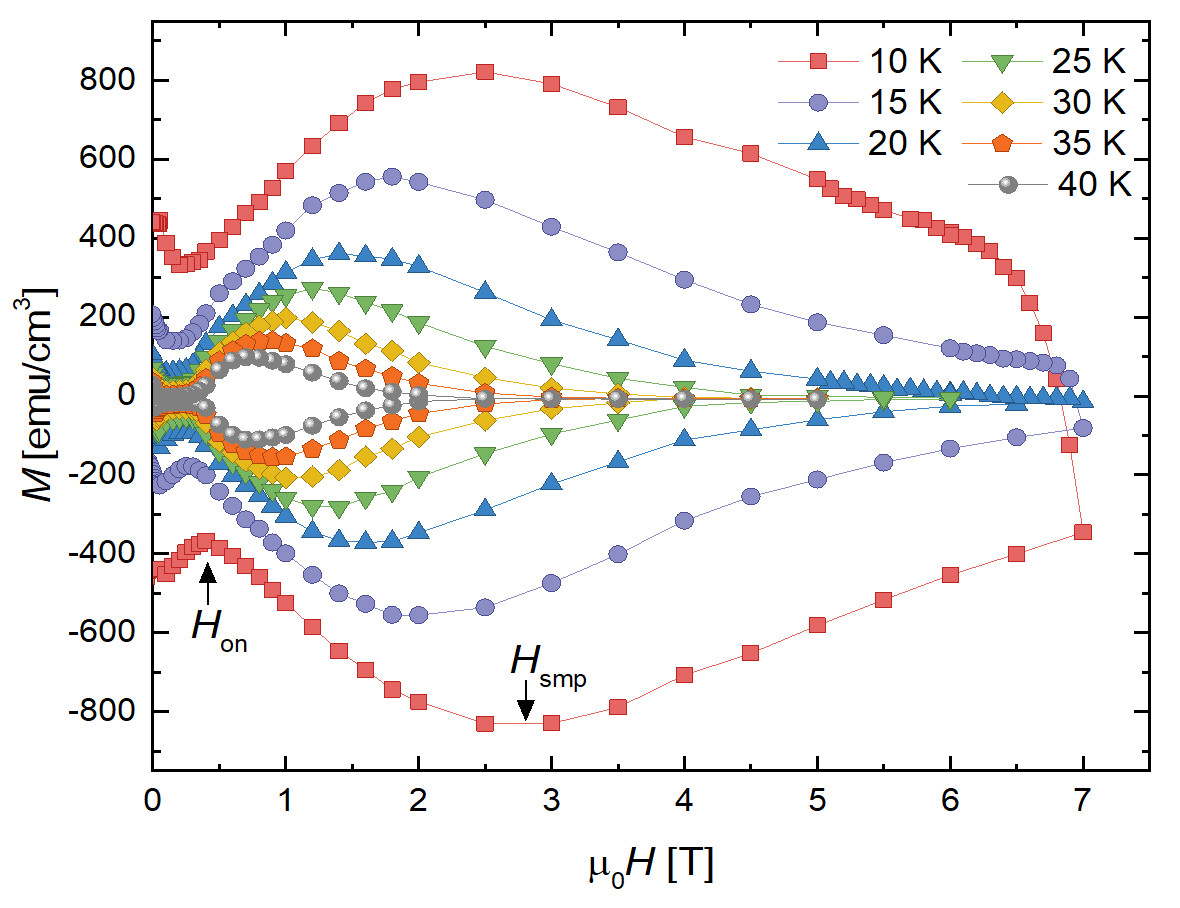}
\caption{\label{Figure2} Isothermal magnetic hysteresis loops $M(H)$ at multiple temperatures for a HgBa$_2$CuO$_{4+\delta}$ single crystal.  Each curve exhibits a low field dip at $H_{on}$ and a second magnetization peak at $H_{smp}$. }
\end{figure}

Isothermal magnetization loops were recorded for $H \parallel c$ and at $T = 5 \text{-} 65 \textnormal{ K}$.  Select curves are displayed in Fig.\ \ref{Figure2}.  In all cases, the field was first swept to -3 or -4 T (not shown) to establish the critical state (full flux penetration throughout the sample).  The lower branch of the loop was subsequently measured as the field was ramped from 0 T to 7 T, and the upper branch was collected as the field was swept back down. All curves exhibit a distinct shape with two conspicuous features: a dip in the magnitude of $M$ near the onset field $H_{on}$ and a second magnetization peak (SMP) at $H_{smp}$.  In general, this shape and the magnitude of the magnetization is indicative of a weak vortex pinning regime at low fields ($H \lesssim H_{on}$) and stronger pinning at higher fields. We observed similar results in measurements of our other Hg1201 crystals. The source of pinning is likely point defects in the Hg-O layer---specifically, oxygen interstitials and mercury vacancies \cite{Wagner1993, Pelloquin1997, Viallet1997}, and this should be the main source of disorder in the bulk that hinders thermal wandering of vortices. 

\begin{figure*}
\includegraphics[width=1\textwidth]{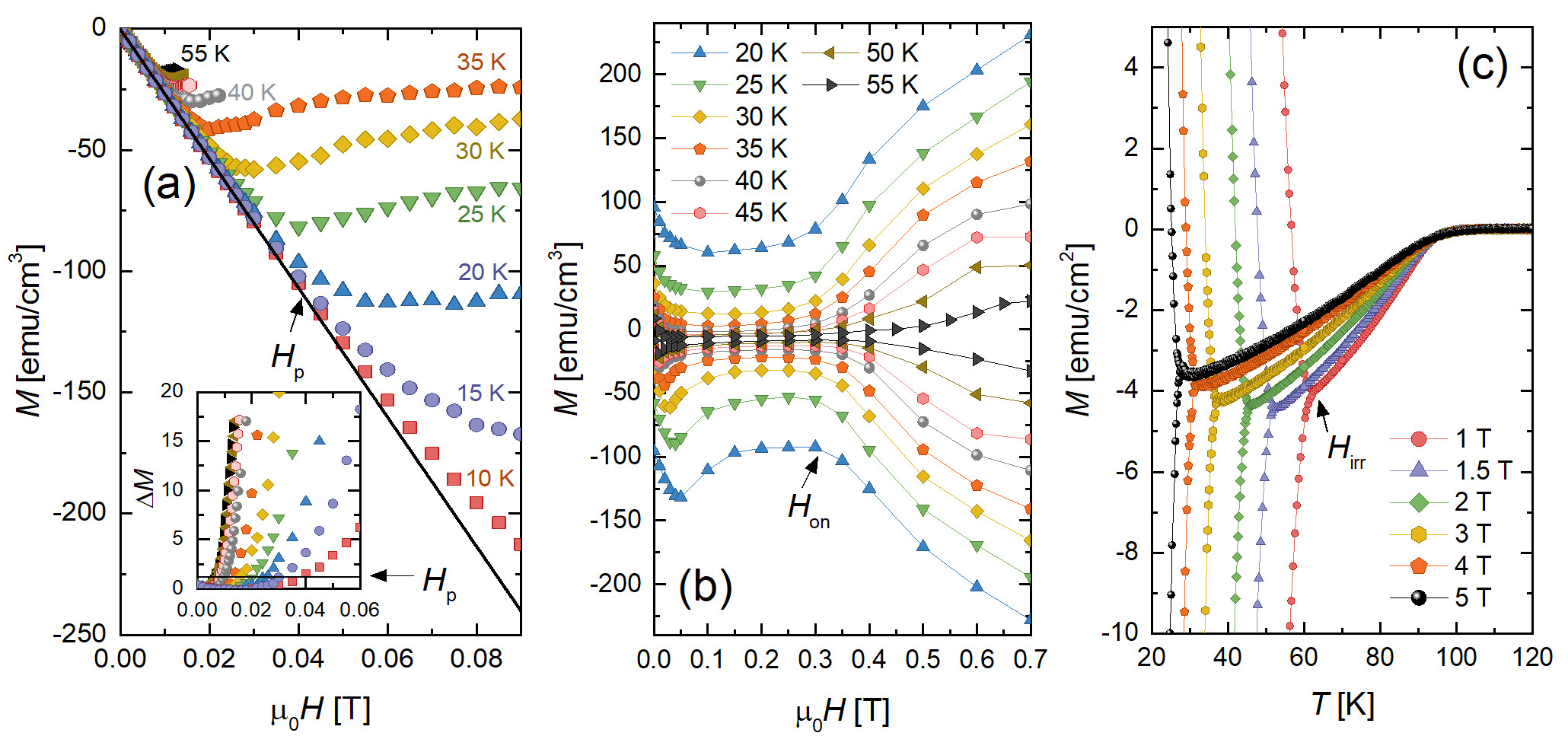}
\vspace{-0.8cm}
\caption{(a) $M$ versus $H$ at low fields after zero field cooling. $H_p$ is the field at which each curve deviates from a linear fit to the Meissner slope (black line). Inset shows the deviation $\Delta M$ from the Meissner slope and the black horizontal line indicates the criterion used for defining $H_p$ (a deviation of a tenth of the standard deviation $\sigma \approx 0.003$ from the linear fit). (b) Magnification of magnetization loops plotted in Fig.\ \ref{Figure2} (collected after full flux penetration) showing a weak pinning regime at low fields in which $M$ is low and weakly sensitive to magnetic field. (c) Temperature dependence of the magnetization showing transition from the irreversible regime to the reversible regime to the normal state.  The upper (lower) branches were collected after the critical state was prepared by sweeping the field to $> \Delta 4 H^*$ above (below)  the indicated fields.  The irreversibility point ($T_{irr}, H_{irr}$) is defined as the point at which the upper and lower branches merge. \label{Figure3}}
\end{figure*}

In high-temperature superconducting crystals, a surface barrier---called the Bean-Livingston (BL) barrier---often plays a significant role in determining the magnetic properties and shaping the $M(H)$ loops \cite{Konczykowski1991, Bean1964, Kopylov1990, Burlachkov1994}.  It originates from competing effects: vortices are repelled from the surface by Meissner shielding currents and attracted by a force arising from the boundary conditions (usually modeled as the attraction between the vortex and a fictitious antivortex image \cite{Burlachkov1993}).  Disappearing at the penetration field $H_p \approx \kappa H_{c1}/\ln \kappa$, the BL barrier impedes vortex entry and exit from the sample in fields less than $H_p > H_{c1}$, where $H_{c1}$ is the lowest field at which flux penetration is thermodynamically favorable. The contribution of the barrier is magnified in materials with high Ginzburg-Landau parameters $\kappa = \lambda/\xi$ and diminished by surface imperfections.  Creep of pancake vortices over the barrier produces the exponential temperature dependence \cite{Burlachkov1994}
\begin{equation} \label{eq:Hp}
H_p(T) \simeq H_c e^{-T/T_0},
\end{equation}
\noindent where $T_0 = \varepsilon_0 d \ln(t/t_0)$, $\varepsilon_0 = \Phi_0^2/4 \pi \mu_0 \lambda_{ab}^2$ is the vortex line energy or tension, $d$ is the spacing between CuO$_2$ planes, $t$ is the time scale of the experiment, $t_0 \sim 10^{-10}-10^{-8} \textnormal{ s}$ relates to the vortex penetration time \cite{Yeshurun1996b}, and the thermodynamic critical field is $H_c=\Phi_0 /2\sqrt{2}\pi\xi_{ab}\lambda_{ab}$.

\begin{figure}[h]
\centering
\includegraphics[width=0.45\textwidth]{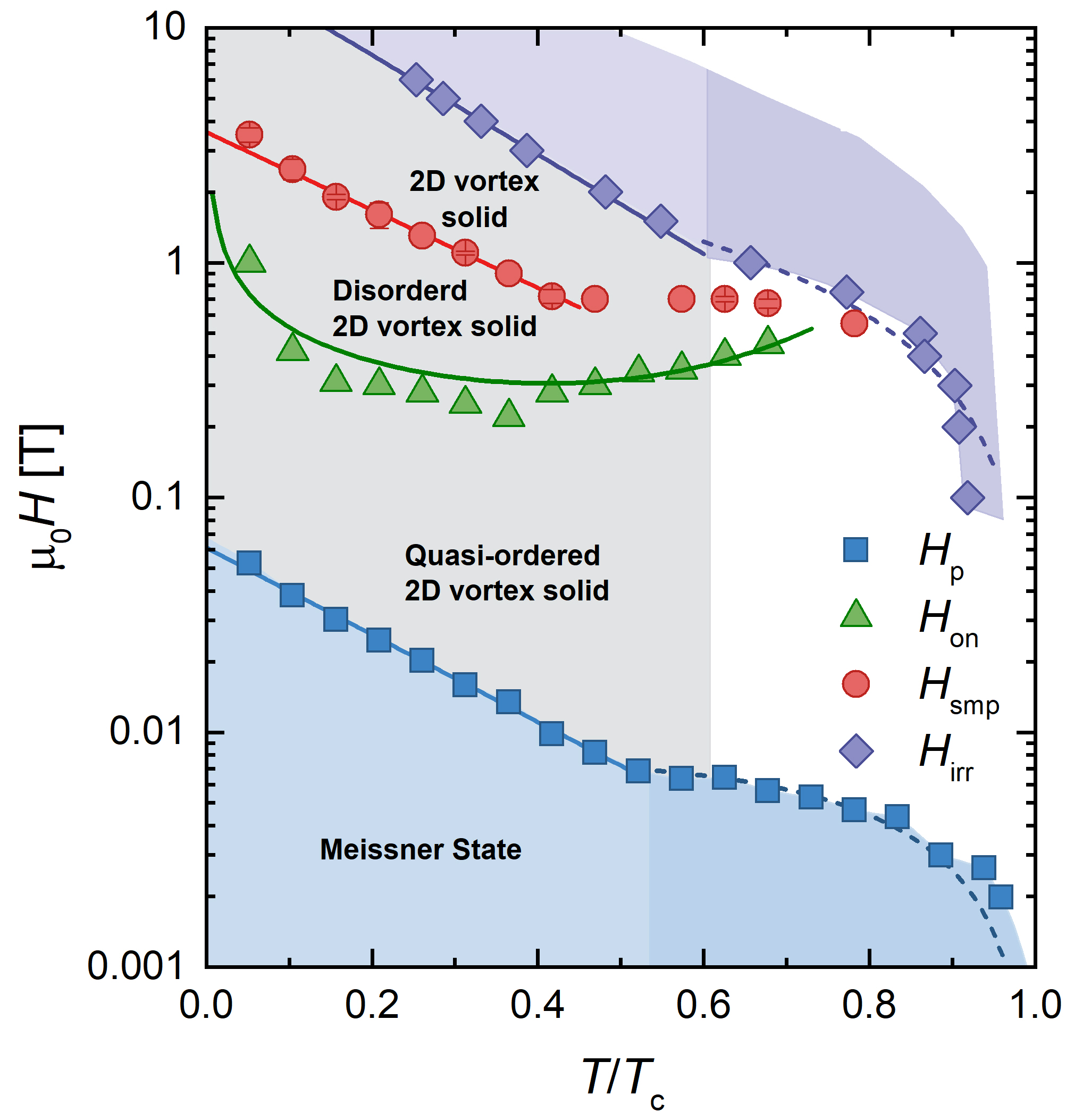}
\vspace{-0.5cm}
\caption{Vortex phase diagram for our Hg1201 crystal, determined by behavior extracted from fits to data shown in Fig.\ \ref{Figure3}.  The solid blue line is a fit to Eq. \eqref{eq:Hp}, while the dashed blue line is a fit to $H_{c1}(T)=[\Phi_0/(4 \pi \lambda_{ab}^2(T)(1-N))]\ln \kappa$.  The solid green curve is a fit of the $H_{on}$ data to Eq. \eqref{eq:Hon}.  The solid red line is a fit of the low temperature second magnetization peak data to $\sim e^{-A T}$ for constant $A$.  Lastly, the solid purple line is a fit to Eq. \eqref{eq:HirrT} and the dashed purple curve is a fit to $(1-T/T_c)^m$.  See text for details.\label{Figure4}}
\end{figure}

To investigate the relevance of surface barriers in our sample, we measured the field at which vortices first penetrate into the sample peripheries $H_p$ by collecting the zero-field cooled $M(H)$ isotherms shown in Fig. \ref{Figure3}(a): we defined $H_p$ as the field at the departure from linearity.  The Figu. \ref{Figure3}(a) inset shows the extraction technique and the phase diagram in Fig.\ \ref{Figure4} contains the resulting temperature dependence.  We find that $H_p(T)$ follows Eq.\ \eqref{eq:Hp} at low temperatures $T/T_c < 0.55$, and a least squares fit produces $T_0 = 21.3 \pm  0.7 \textnormal{ K}$ and $H_c = 0.06 \textnormal{ T}$.  The experimentally extracted $T_0$ is reasonably close to the estimate $T_0 = 26 \textnormal{ K}$, calculated assuming $t \sim 100 \textnormal{ s}$, $t_0 = 10^{-10} \textnormal{ s}$, $d \approx 9.5 \textnormal{ \AA}$ \cite{Bras1996}, and $\lambda_{ab} \approx 162 \textnormal{ nm}$ \cite{Villard1999}.  

To assess the accuracy of the extracted $H_c$, we must first account for demagnetizing effects by multiplying its value by $1/(1-N)$, for $N$ is the effective demagnetizing factor (see \footnote{While expressing the field enhancement in terms of a demagnetizing factor formally applies to samples with elliptic cross-sections only, it is often used for slabs and disks as well. For samples with rectangular cross-section (width $w$ and thickness $\delta$, $w \gg \delta$) the field-penetration is retarded by a geometrical barrier\cite{Zeldov1994, Willa2014}. This barrier is associated with a parametrically [by a factor $(\delta/w)^{1/2}$] lower field enhancement at the sample edge as compared to an elliptic slab with the same dimensions. The field of first penetration scales as $H_{p} \propto (w/\delta)^{1/2} H_{c1}$ and is in agreement with the numerical values found by Brandt \cite{Brandt2001}} for comment on demagnetizing factors in slabs). Because analytic expressions for $N$ are only known for infinite slabs and disks, we can either follow common procedure and approximate the sample as a disk or, more accurately, use numerically calculated effective demagnetizing factors for rectangular prisms. In the former case, we calculate $N \approx 0.65$ using expressions derived by Brandt \cite{Hcl1965}. In the latter, numerical calculations from Ref. \cite{Pardo2004} produce $N \approx 0.75$ such that $H_c = 0.24 \textnormal{ T}$.  This yields a coherence length $\xi_{ab}(0) \sim 1.5 \textnormal{ nm}$, similar to the value of $\xi_{ab}(0) \sim 2.0 \pm 0.4 \textnormal{ nm}$ measured by Hofer et al.  \cite{Hofer1998}.

Agreement of our $H_p(T<0.55 T_c)$ data with Eq.\ \eqref{eq:Hp} indicates that, at low $T$, vortices enter as pancakes and are thermally activated over the BL barrier.  At $T/T_c \sim 0.55 $, the temperature dependence of $H_p$ abruptly changes suggestive of a different mechanism for vortex penetration at higher temperatures.  Similar crossovers have been observed in other layered superconductors \cite{Chowdhury2004, Chowdhury2003, Kopylov1990, Motohira1991, Zeldov1995, Niderost1998} around $T/T_c \sim 0.5$.  We consider two possible explanations: we are observing a crossover from creep of pancakes to creep of half-loops (3D vortex lines) over the barrier or $H_p$ becomes less than $H_{c1}$.  In the former scenario, Josephson interaction between pancakes becomes significant and vortices enter via creep of half-loop excitations.  This would produce a temperature dependence $H_p \propto (T_c - T)^{3/2}/T$ \cite{Burlachkov1994, Niderost1998}, which fails to fit our data.  However, in the latter case, we would expect $H_p (T) \approx (1-N)H_{c1}(T)=[\Phi_0/(4 \pi \lambda_{ab}^2(T))]\ln \kappa$ for $T$ above the crossover. Considering the two-fluid approximation $\lambda_{ab}(T) = \lambda_{ab}(0) [1-(T/T_c)^4]^{-1/2}$ and a $T$-independent $\kappa$, this expression indeed fits the data for $300 \textnormal{ Oe}$ (see Fig. \ref{Figure4}).  Considering this value for $H_{c1}$ and a coherence length $\xi_{ab} \approx 2 \textnormal{ nm}$, we find $\lambda_{ab}(0) = 154 \textnormal{ nm}$, which is comparable with published data \cite{Villard1999, Hofer1998}.  We therefore conclude that the crossover in the temperature dependence of $H_p$ originates from a reduction in the BL surface barrier below $H_{c1}$ at high temperatures.
 
For fields above $H_p$, but below $H_{smp}$, $|M|$ dips to an ill-defined minimum and increases again at $H_{on}$.  Figure \ref{Figure3}(b) magnifies this low-field plateau and $H_{on}(T)$ is plotted in Fig. \ref{Figure4}. Previous studies have related $H_{on}$ to a transition between an ordered vortex lattice at low fields and an entangled lattice created by point disorder at higher fields, tuned by competition between thermal, pinning, and elastic energies. For example, FeSe$_{1-x}$Te$_x$ single crystals showed evidence of a Bragg glass (quasi-ordered vortex solid) below $H_{on}$ \cite{Miu2012} and a presumed disordered vortex solid above $H_{on}$.  Additionally, YBa$_2$Cu$_3$O$_{7-\delta}$, Nd$_{1.85}$Ce$_{0.15}$CuO$_{4-\delta}$, and Bi$_2$Sr$_2$CaCu$_2$O$_{8+\delta}$ all demonstrate disorder induced phase transitions that show a signature in the $M(H)$ loops \cite{Giller1999}.

With increasing field, the elastic energy $E_{el}$ decreases and when it becomes comparable to the pinning energy $E_{pin}$, the lattice evolves into a disordered state at $H_{k}(T)$. Generally speaking, vortex pinning is caused by spatial variations of the Ginzburg-Landau order parameter, capturing disorder in either $T_c$ or in the charge carrier mean free path $\ell$ near defects \cite{Griessen1994}.  This is commonly referred to as ``$\delta T_c$ pinning" and ``$\delta \ell$ pinning", respectively, and it is thought that the temperature dependence of $H_k$ can identify the responsible pinning mechanism \cite{Giller1999, Chowdhury2003}. In the case of $\delta T_c$ pinning, 
\begin{equation}\label{eq:Hon2}
    H_{on}(T)=H_{on}(0)[1-(T/T_c)^4]^{3/2},
\end{equation}
whereas for $\delta \ell$ pinning,
\begin{equation}\label{eq:Hon1}
H_{on}(T)=H_{on}(0)[1-(T/T_c)^4]^{-1/2}.
\end{equation}
 
Eqs.\ \eqref{eq:Hon2} and \eqref{eq:Hon1} fail to capture the non-monotonic behavior demonstrated in our measurements.  On the contrary, if strong thermal fluctuations cause the amplitude of the vortex line to be comparable to $\xi$ at a depinning temperature $T_{dp}$, then the temperature dependence of  $H_{on}$ is expected to be \cite{Chowdhury2003, Giller1999, Erta1996}
\begin{equation} \label{eq:Hon}
H_{on}(T) = H_{on}(0) \left[ \frac{T_{dp}}{T} \frac{\displaystyle e^{(T/T_{dp})^3-1}}{1 - ( T / T_c )^4}\right]^{1/2}.
\end{equation}
\noindent As can be seen from the fit (green curve) in Fig. \ref{Figure4}, Eq. \eqref{eq:Hon} not only captures the non-monotonicity, but also fits the data quite well for $T_{dp} = 59.5 \pm 6.4 \textnormal{ K}$ and $H_{on}(0)=0.35 \pm 0.03 \textnormal{ T}$. Note that $T_{dp}$ was found to be 66 K in a YBCO crystal \cite{Giller1999} and 32 K for Tl$_2$Ba$_2$CaCu$_2$O$_8$ \cite{Chowdhury2003}.  

In applied magnetic fields above $H_{on}$, the magnetization apexes at the second magnetization peak $H_{smp}$. Second magnetization peaks have been reported in studies of most classes of superconductors, including low-$T_c$ \cite{Banerjee2000, Peng1994}, iron-based \cite{Miu2012, Fang2011, Zhou2016a, Salem-Sugui2010, Pramanik2011a, Shen2010}, and highly anisotropic \cite{Chowdhury2003, Konczykowski2000} materials, as well as YBa$_2$Cu$_3$O$_{7-\delta}$ (YBCO) single crystals \cite{Boudissa2006a}.  In fact, this peak has also been observed in a few previous studies \cite{Daignere2000, Villard1999, Pissas1999, Pissas1998, Stamopoulos2002} of Hg1201 single crystals grown by two research groups \cite{Pelloquin1997, Pissas1998}, though the peak is far more pronounced in our samples.  This feature is typically either attributed to a crossover between vortex pinning regimes or a structural phase transition in the vortex lattice.  In Sec. IIIF, we will revisit the discussion of the second magnetization peak because creep measurements are requisite to evaluate possible origins of the SMP.

At sufficiently high fields, the loops close as the system transitions into the reversible regime at the irreversibility field $H_{irr}$.  Instead of extracting $H_{irr}$ from the magnetization loops, we extract it from isomagnetic $M(T)$ sweeps.  This is more precise than measurements involving sweeping the field: temperature sweeps tend to induce less noise than field sweeps and, at the transition, the upper and lower branches of $M(T)$ not only converge, but also exhibit a sharp change in slope.  Figure \ref{Figure3}(c) contains select $M(T)$ datasets showing the extraction technique and the resulting irreversibility line is shown in Fig.\ \ref{Figure4}.

We now consider the possible origin of irreversibility.  In type-II superconductors, either vortex pinning or barriers (geometric or surface) can cause $M(H,T)$ to be irreversible \cite{Brandt2001}.  Such barriers can indeed control the position of the irreversibility line even in samples in which the width of the magnetization loop is determined by significant bulk pinning \cite{Burlachkov1994, Brandt2001, Zeldov1995}.  In fact, here we find that, at low temperatures $T/T_c \lesssim 0.6$, $H_{irr}$ has the same functional form for the dependence on temperature as the penetration field, 
\begin{equation} \label{eq:HirrT}
H_{irr}(T) \propto e^{-T/T_0}.
\end{equation}
The fit of our $H_{irr}$ data for $T/T_c < 0.6$ to Eq. \eqref{eq:HirrT} is shown in Fig.\ \ref{Figure4}, which yields $T_0 = 19.7 \pm 0.6 \textnormal{ K}$, produced by a least squares fit.  Notice that $T_0$ is close to the value $T_0 \approx 21 \textnormal{ K}$, extracted in the fit of our $H_p$ data to Eq. \eqref{eq:Hp} and identical to the value ($T_0 = 19.7 \pm 0.4 \textnormal{ K}$) extracted in another study on Hg1201 single crystals \cite{Pissas1998}.  This dependency suggests that the irreversibility line in our Hg1201 crystal is controlled by surface or geometric barriers for $T/T_c < 0.6$, and that the phase transition across the irreversibility line is driven by thermal activation of pancake vortices \cite{Burlachkov1994}.  The irreversibility line may also coincide with the melting line $H_m$. According to a model for superconductors having moderate anisotropy \cite{Schilling1993}, $H_m$ is expected to decrease with $T$ as $H_m \propto e^{-\beta/T}$ (for constant $\beta$) at low temperatures.

At higher temperatures $T>T^*$, the shape of the irreversibility line changes, suggesting a change in the mechanism controlling irreversibility.  Similar trends in $H_{irr}(T)$ have been found in grain-aligned Hg1201 samples \cite{Lewis1995} and single crystals \cite{Pissas1998}.  However, in the former, 2D dynamics were observed over the entire measurement range and, in the latter, over most of the range ($T \leq 89 \textnormal{ K}$).  
Our high temperature data (in clean crystals) fits $H_{irr} \propto (1-T/T_c)^m$, where a least squares fit yields $m =1.05$.  This dependence is roughly consistent with the expectation for the melting line for 3D-like vortex fluctuations \cite{Brandt1989, Houghton1989, Blatter1994, Sasagawa1998, Schilling1993, Tian1997}.  Hence, the irreversibility line may coincide with melting over the entire temperature range and the change in behavior of $H_{irr}$ at high temperatures is indicative of a dimensional crossover from dynamics driven by 2D pancakes to the 3D lines.  We should, however, note that fits are not precise enough to prove exact coincidence between irreversibility and melting, and the melting line could simply lie nearby the irreversibility line.  
 
The exponentially decaying temperature dependence of both $H_p$ and $H_{irr}$ over roughly half of the phase diagram suggests a strong contribution of the surface barrier to the vortex dynamics.  This raises the question of the relative contributions of bulk pinning. If bulk pinning were completely disregarded, an imbalance in the rate of flux entry and exit from the sample would arise and produces asymmetric $M(H)$ loops with negligible magnetization in the lower branch compared to the upper (or vice versa) \cite{Burlachkov1993}.  However, the upper and lower branches of our magnetization curves are roughly symmetric, adhering to expectations of the Bean model for bulk pinning.  Hence, we conclude that the bulk is the dominant pinning source in our Hg1201 crystal.
Compiling the aforementioned results, Fig.\ \ref{Figure4} shows the resulting phase diagram on a semilog plot.  In the following sections, we use magnetic relaxation measurements to learn more about the nature of vortex dynamics in the gray region of Fig.\ \ref{Figure4}. The following sections present our main result -- a more detailed understanding of vortex behavior derived from extensive vortex creep measurements.

\subsection{C. Vortex creep as a function of magnetic field}
The disorder landscape defines potential energy wells in which vortices will preferentially localize to reduce their core energies by a pinning energy $U_0$.  An applied or induced current tilts this energy landscape.  This reduces the energy barrier that a pinned vortex must surmount to escape from a well to a current-dependent value $U(J)$.  The time required for thermal activation over such a barrier can be approximated by the Arrhenius form 
\begin{equation} \label{eq:creeptime}
t=t_0 e^{U(J)/k_BT}.
\end{equation}
At low temperatures ($T \ll T_c$) and fields, the simple linear relationship $U(J)=U_0 (1-J / J_{c0})$ proposed in the Anderson-Kim model \cite{Anderson1964, Blatter1994b} is often accurate.  However, because this model neglects vortex elasticity and vortex-vortex interactions, its relevance is often further limited to the early stages of the relaxation process ($J \lesssim J_{c0}$). In the later stages $J/J_{c0} \ll 1$, collective creep theories, which consider vortex elasticity, predict an inverse power law form for the energy barrier $U(J)=U_0[(J_{c0} / J)^\mu]$. Here, the glassy exponent $\mu$ is sensitive to the size of the vortex bundle that hops during the creep process and its dimensionality.  To capture behavior for a broad range of $J$, we invoke a commonly used relationship that interpolates between the two regimes 
\begin{equation}\label{eq:Uinterpolation}
U(J)=\frac{U_0}{\mu} [(J_{c0}/J)^\mu -1], 
\end{equation}
\noindent where $\mu =-1$ recovers the Anderson-Kim result.  It is now straightforward to combine Eqs.\ \eqref{eq:creeptime} and \eqref{eq:Uinterpolation} to determine the expected decay in the persistent current over time $J(t)$ and subsequently the vortex creep rate $S$: 
\begin{equation}\label{eq:Jtdecay}
J(t)=J_{c0} \left[1+\frac{\mu k_B T}{U_0} \ln (t/t_0)\right]^{-1/\mu}
\end{equation}
and 
\begin{equation}\label{eq:ST}
S \equiv \left| \frac{d \ln J}{d \ln t} \right| = \frac{k_B T}{U_0+\mu k_B T \ln (t/t_0)}.
\end{equation}
Creep measurements are a useful tool for determining the size of the energy barrier and its dependence on current, field, and temperature.  Such measurements further probe the vortex state, revealing the existence of glassy behavior, collective creep regimes, or plastic flow.   This is because, as evident in Eq.\ \eqref{eq:ST}, creep provides access to both $U_0$ and $\mu$.  Table \ref{TableMu} summarizes expected values of $\mu$ for collective creep of 3D flux lines and 2D pancake vortices.

\begin{table}[H]  
\centering
 \caption{Exponents $\mu$ predicted by collective creep theory \cite{Blatter1994b, Vinokur1995}. Exponents depend on the dimension and size of element that hops due to thermal activation. Specifically, $\mu$ depends on whether it is a single vortex or a vortex bundle of lateral dimension $R_c$ smaller than (small bundle), comparable to (medium bundle), or larger than (large bundle) the penetration depth $\lambda_{ab}$. \label{TableMu}}
 \begin{tabular}{l l l}
 \hline
 \multicolumn{2}{c}{} \\[-11pt]
 \hline
 Dimension & 	Single vortex or bundle size     &	$\mu$ \\
 \hline
3D  &	Single vortex    &	1/7 \\
3D  &	Small vortex bundles	    & 3/2, 5/2 \\
3D  &   Large vortex bundles    &	7/9 \\
2D  &   Small vortex bundles    &	7/4 \\
2D  &   Medium vortex bundles    & 	13/16 \\
2D  &   Large vortex bundles    & 1/2 \\
\hline
\multicolumn{2}{c}{} \\[-11pt]
\hline
\end{tabular}
 \end{table}
 
 \begin{figure}
 \centering
\includegraphics[width=0.45\textwidth]{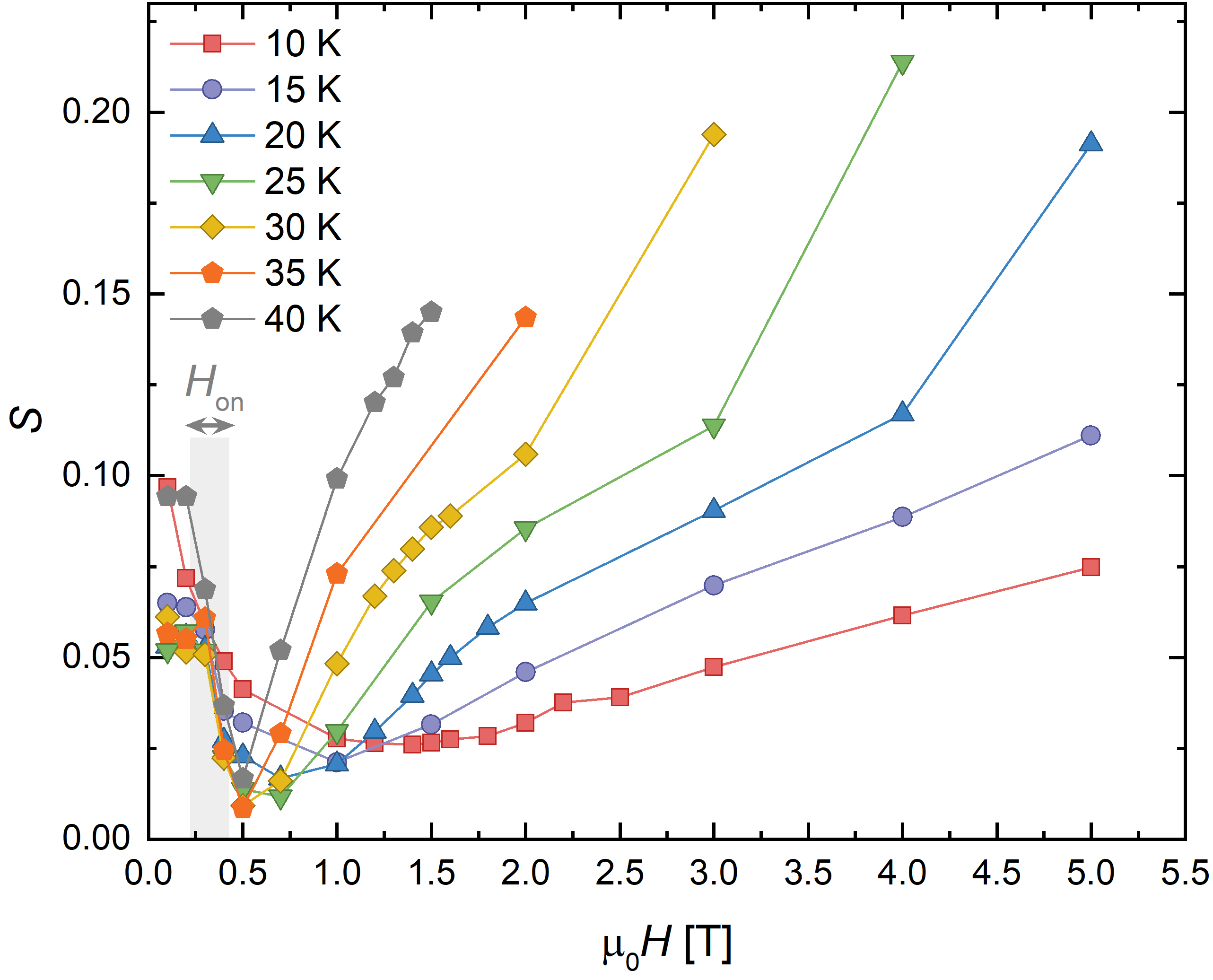}
\caption{Field dependence of the magnetic relaxation rate at temperatures 15-45 K.  Non-monotonicity is suggestive of different dynamics at low versus high magnetic fields.\label{FigureSH}}
\end{figure}

\begin{figure*}
\includegraphics[width=0.8\textwidth]{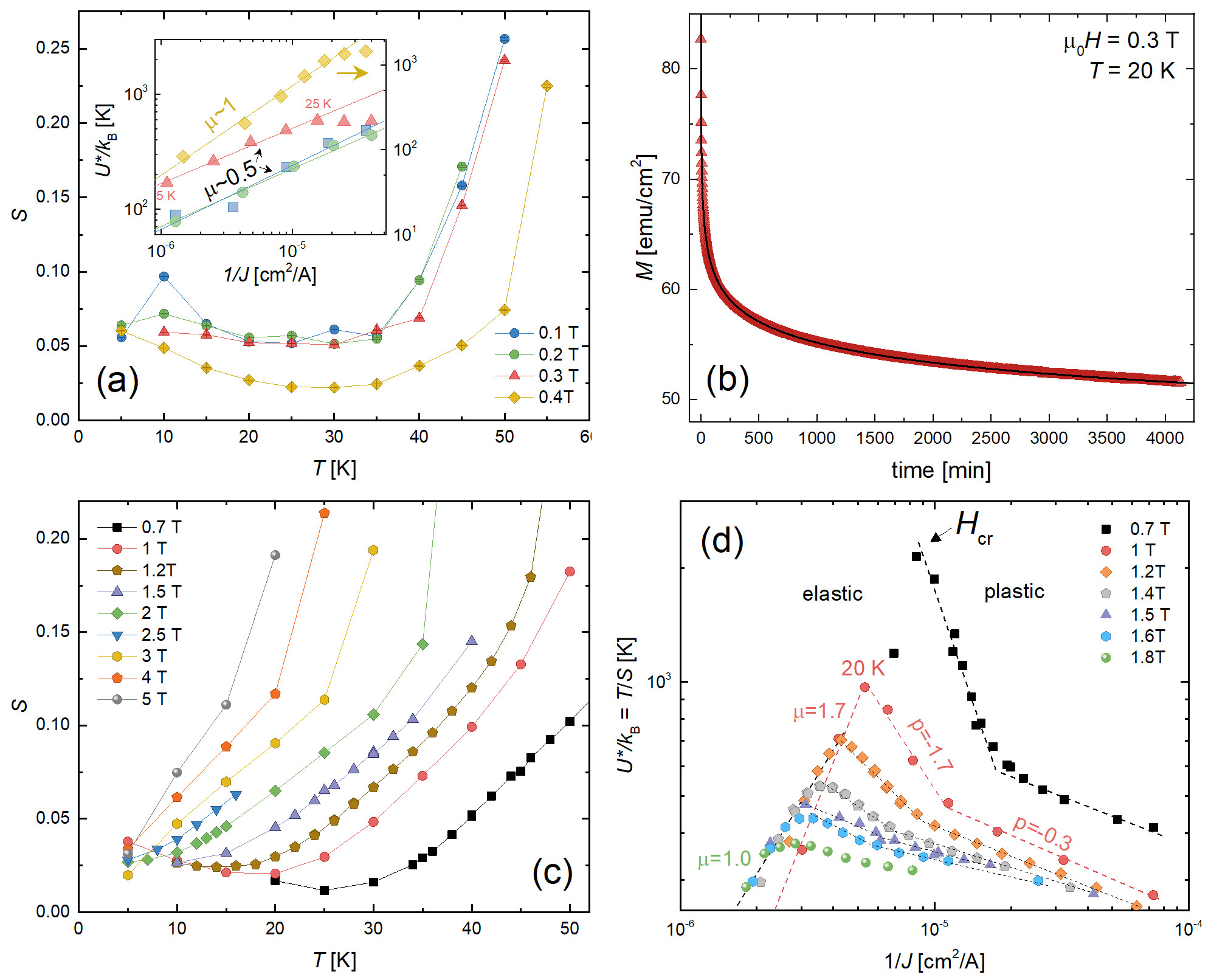}
\caption{Temperature dependence of the vortex creep rate in applied magnetic fields of (a) 0.1 - 0.4 T and (c) 0.7 - 5 T. The inset to (a) shows the energy scale $U^* \equiv k_B T/S$ versus $1/J$. The lines are linear fits to the data for temperatures 5 - 25 K, and the slopes yield the glassy exponents $\mu  \approx 1$ at 0.4 T and $\mu \approx 0.5$ for 0.1 - 0.3 T.  (b) Magnetization ($M$) collected every $\sim$15 s for 67 hours at 20 K and 0.3 T.  The black curve is a fit to Eq.\ \eqref{eq:Jtdecay} using $\mu = 0.5$, where $J(t) \propto M(t)$.  (d) Energy scale $U^*$ plotted against $1/J$ for applied field 0.7 - 1.8 T.  The lines are linear fits, and the change from a positive to negative slope suggests a crossover from elastic vortex dynamics to plastic flow at $H_{cr}$.  The dashed lines show examples of how the glassy exponents $\mu$, displayed in the phase diagram in Fig. \ref{Figure8}, were extracted. \label{Figure6}} 
\end{figure*}       

To shed light on the Hg1201 phase diagram, we measured creep rates in a wide range of temperatures (5-60 K) and magnetic fields (0.1 - 5 T) using standard methods \cite{Yeshurun1996b}, summarized here.  We first establish the critical state by sweeping the field $4H^*$ above the field at which creep will be measured $H$, where $H^*$ is the minimum field at which magnetic flux will fully penetrate the sample.  Second, the field is swept to $H$, such that the magnetization $M(H)$ coincides with  its value on the upper branch of a magnetization loop. [If the magnitude of the initial field sweep were not sufficiently high, $M(H)$ would instead fall inside the loop, vortices would not fully penetrate the entire sample and the previously discussed models would be inapplicable.]  Third, the magnetization $M(t) \propto J(t)$ is subsequently recorded every $\sim 15$ s for an hour.  We also briefly measure $M(t)$ in the lower branch to determine the background arising from the sample holder, subtract this, and adjust the time to account for the difference between the initial application of the field and the first measurement (maximize correlation coefficient).  Lastly, the normalized creep rate $S(T,H)$ is extracted from the slope of a linear fit to $\ln M - \ln t$.

Figure \ref{FigureSH} shows the field dependence of the creep rate.  In low fields, $S$ decreases as $H$ increases, a trend that reverse above $\sim$0.5 T. This change in behavior may be related to a different source of vortex pinning at low than at high fields and roughly coincides with the low-field change in shape of the $M(H)$ loops around $H_{on}\sim$ 0.5 T.  Because of this, in the following section, we will separately analyze low-field and high-field measurements.  We will first present $S(T)$ and an analysis of the vortex state in the low-field regime, and then proceed to analyze the high-field regime.

\subsection{D. Glassy vortex dynamics and elastic-to-plastic crossovers}
\label{sec:EtoPcrossovers}
To study the dynamics in the low-field weak pinning regime, exemplified in Fig.\ \ref{Figure3}(b), we measured vortex creep for $\mu_0H < 0.5 \textnormal{ T}$, shown in the main panel of Fig.\ \ref{Figure6}(a).    The creep rates in fields of 0.1 - 0.3 T are similar over the entire temperature range, plateauing at $S \sim 0.06$ for $T < 40 \textnormal{ K}$ then sharply rising at higher temperatures.   Such behavior is akin to $S(T)$ in YBCO samples, which typically exhibit a plateau around $S \sim 0.02 - 0.035$ \cite{Malozemoff1990b, Yeshurun1996b, Kwok2016}.  In YBCO, the plateau appears because $U_0 \ll \mu T \ln(t/t_0)$ such that $S \sim [\mu k_B \ln (t/t_0)]^{-1}$ becomes $T$-independent.  It is often associated with glassy vortex dynamics because $\mu \approx 1$ considering $S \sim 0.035$ and $\ln (t/t_0) \approx 27$  for a typical measurement window of $t \sim$1 hour \cite{Yeshurun1996b, Kwok2016}.  

Similarly, for our Hg1201 sample, if $U_0 \ll \mu k_B T \ln(t/t_0)$ were true, the $S \sim 0.05$ plateau would yield $\mu \sim 0.6$.  However, in our sample, we do not yet know the comparative magnitudes of $U_0$ and $\mu k_B T \ln (t/t_0)$.  To extract $\mu$ without the need for assumptions regarding $U_0$, it is common practice to define an experimentally accessible energy scale $U^* \equiv k_B T/S$.  From Eq.\ \eqref{eq:ST}, we see that $U^* = U_0 + \mu k_B T \ln (t/t_0)$ and, combined with Eq.\ \eqref{eq:Jtdecay}, find that 
\begin{equation}\label{eq:Ustar}
U^*\equiv \frac{k_B T}{S} = U_0\left(\frac{J_{c0}}{J}\right)^\mu.
\end{equation}


Hence, $\mu$ can be directly obtained from the slope of $U^*$ versus $1/J$ on a $\log-\log$ plot. As shown in the Fig.\ \ref{Figure6}(a) inset, $\mu = 0.5$ for fields of 0.1 - 0.3 T.  We reinforce this result with a complementary 67 hour long relaxation study shown in Fig.\ \ref{Figure6}(b) and fitting the resulting $M(t)$ to the interpolation formula Eq.\ \eqref{eq:Jtdecay}.  For free parameters $J_{c0}$, $U_0$, and $\mu$, the best fit again produces $\mu = 0.5$, which is expected for collective creep of large bundles of 2D pancake vortices (see Table \ref{TableMu}) \cite{Vinokur1995}.

The presence of large bundles in these small fields is suggestive of a clean pinning landscape in which long-range $1/r$ vortex-vortex interactions are only weakly perturbed by vortex-defect interactions.  Furthermore, this result is consistent with the evidence from $H_{irr}(T)$ (shown in Fig.\ \ref{Figure4}) of a 2D vortex state over a wide low temperature $T/T_c \ll 0.6$ region of the phase diagram.   We have now ascertained that, though the plateau in $S(T)$ appears at a higher $S$ than in YBCO, it again correlates with glassiness.

At 0.4 T, $S(T)$ is non-monotonic, reaching a local minimum around 30 K (see Fig. \ref{Figure6}(a)). 
As shown in the Fig. \ref{Figure6}(a) inset, we extract $\mu \approx 1$, which is close to the $\mu = 13/16$ expectation for creep of medium bundles of pancake vortices (see Table \ref{TableMu}). So, the system transitions from creep of large bundles at low fields $\mu_0 H < 0.4 T$ to medium bundles at 0.4 T.  This change in $\mu$ occurs roughly around $H_{on}$ (compare to Fig. \ref{Figure3}(b)) and the minimum in $S(H)$ (compare to Fig. \ref{FigureSH}).  In many systems, the bundle size increases with increasing $H$ \cite{Blatter1994b}.  Hence, this scenario is not standard, but is consistent with our suspected mechanism for $H_{on}$: as $H$ increases, the strength of pinning suddenly increases around $H_{on}$ causing the lattice to become more entangled, the bundle size to decrease, and we see both $J_c$ and $\mu$ increase.

Collective creep theory only considers elastic deformations of the vortex lattice and neglects dislocations.  At high temperatures and/or fields, the elastic pinning barrier becomes quite high and plastic deformations of the vortex lattice can become more energetically favorable. Plastic creep \cite{Kierfeld2000} involves the motion of a channel of vortices constrained between two edge dislocations of opposite sign (dislocation pairs) and requires surmounting a diverging plastic barrier $U_{pl} \sim J^{-\mu}$ for small driving force $J \ll J_{c0}$.  It manifests as a negatively sloped region in a $U(1/J)$ plot: in Eq.\ \ref{eq:Ustar}, $\mu < 0$ is conventionally represented using the notation $p$, such that $U^* = U_0(J_{c0}/J)^p$ in the plastic regime.

\begin{figure}
\centering
\includegraphics[width=0.45\textwidth]{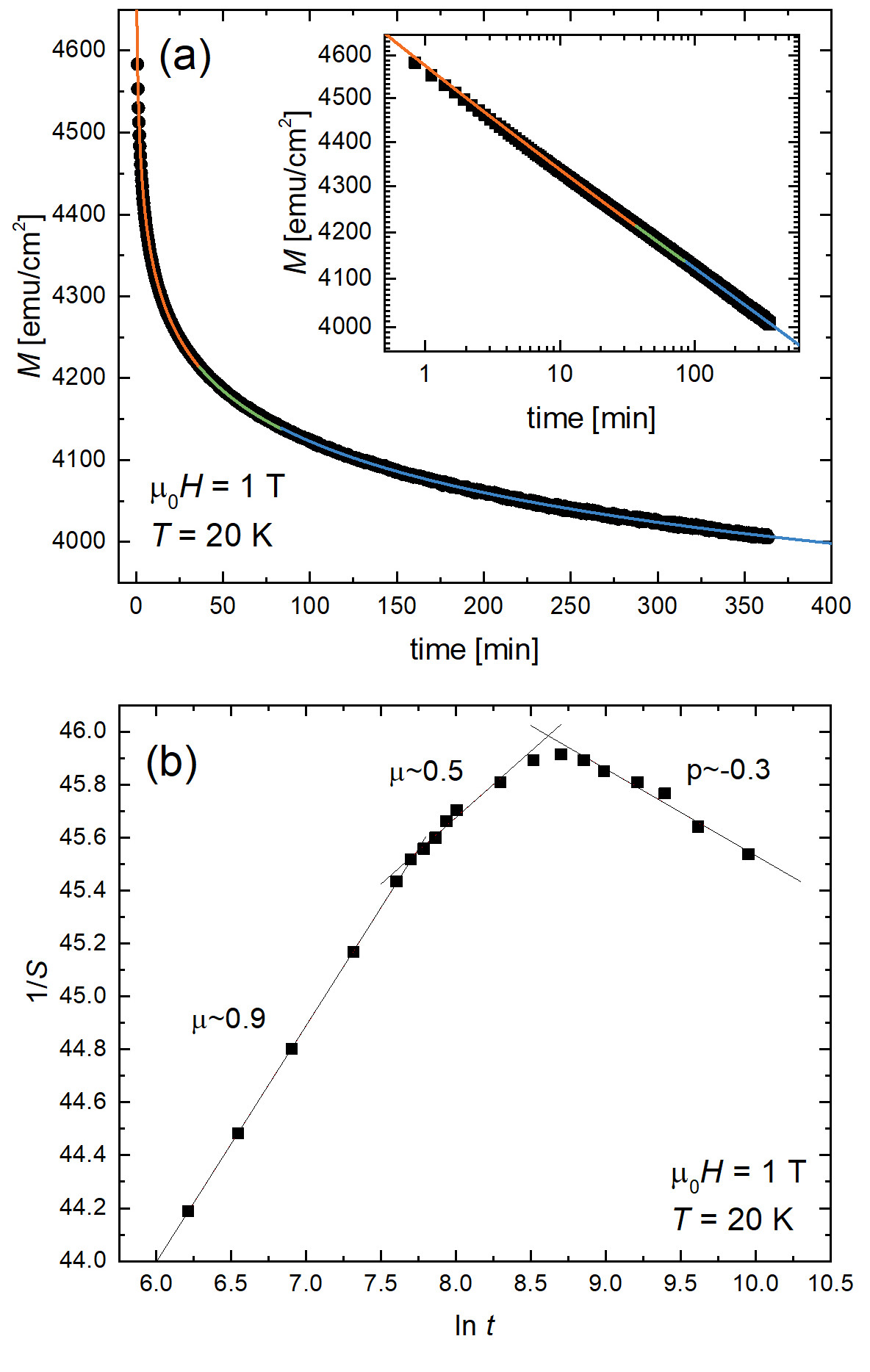}
 \caption{$1/S$ versus $\ln t$ determined from the time-dependent magnetization $M(t)$ data shown in the inset.  $M(t)$ was collected every $\sim$15 s for 6 hours at 20 K and 1 T, at the elastic-to-plastic crossover.  The data points in the main panel represent $S$ extracted from subsets of the data from times 0 to $t$.  Dashed lines are linear fits.  The red, green, and blue curves in the inset are fits to Eq.\ \eqref{eq:Jtdecay}, where $J(t) \propto M(t)$, using $\mu =0.9$, $\mu =0.5$, and $\mu \rightarrow p =-0.3$, respectively.  \label{FigureCrossover}}
\end{figure}

Figure \ref{Figure6}(c) displays creep rates at fields $H > 0.5 \textnormal{ T}$.  Representing the data as $U^*(1/J)$, plotted in Fig.\ \ref{Figure6}(d), the slopes exhibit a distinct sign change, revealing elastic ($\mu > 0$) to plastic ($\mu <0 \rightarrow p$) crossovers for $H \leq 2 \textnormal{T }$. Figure\ \ref{Figure6}(d) displays the 1 T data for fields of 0.7 -- 1.8 T.  From the data, we extract the exponents $\mu$ displayed in the vortex phase diagram show in Fig. \ref{Figure8}(a).  We see, for example, that at $1 \textnormal{ T}$ the sample hosts creeping small bundles of pancake vortices in the elastic regime $T < 20$ \textnormal{ K}.  To investigate the dynamics at the crossover temperature $T_{cr} = 20 \textnormal{ K}$, we perform a 6-hour measurement of $M(t)$ that is plotted in the Fig.\ \ref{FigureCrossover} and fit the data to Eq.\ \eqref{eq:Jtdecay}. To reduce the number of free parameters ($J_{c0}$ and $U_0$), we obtain $\mu$ from the slope of $1/S$ plotted against $\ln t$, see Eq.\ \eqref{eq:ST} and the Fig.\ \ref{FigureCrossover} inset. Clearly, the early stages of relaxation is glassy. The bundle size evolves over time, manifesting as a change in $\mu$. In the latter stages, we observe a transition to plastic flow.

\subsection{F. Second magnetization peak and vortex phase diagram}
Elastic-to-plastic crossovers are considered the cause of the second magnetization peak in many superconductors \cite{Miu2012, Zhou2016a, Shen2010, Salem-Sugui2010}. However, the origin of the SMP in Hg1201 is controversial. Daign\`ere \textit{et al.} \cite{Daignere2001, Daignere2000} found no correlation between the SMP and elastic-to-plastic crossovers in Hg1201 single crystals, and concluded that the SMP merely arises from competition between an increase in $J_c$ and decrease in pinning energy with increasing magnetic field.  On the contrary, Pissas \textit{et al.} \cite{Pissas1999} showed that though $H_{cr} < H_{smp}$, a correlation between the two fields does indeed exist, therefore the peak is possibly associated with collective-to-plastic transitions.  That study reconciled the lack of coincidence between the two fields as caused by very fast creep at low fields and slower creep at high fields. To understand their reasoning, it is important to note that magnetization measurements are not collected instantaneously with the application of a magnetic field.  That is, there is a $10-100 \textnormal{ s}$ lag between establishing the field and measuring $M$ due to the time required for setting the magnet in persistent mode and translating the sample through the magnetometer SQUID detection coils.  Consequently, by the time $M$ is recorded during magnetization loop measurements, $J$ is much less than $J_{c0}$ at low fields where creep is fast and closer to $J_{c0}$ at higher fields where creep is slow.  This idea was further supported by a demonstration that measuring the loop faster shifts $H_{smp}$ to lower fields, towards $H_{cr}$ \cite{Pissas1999}.

\begin{figure}
\centering
\includegraphics[width=0.4\textwidth]{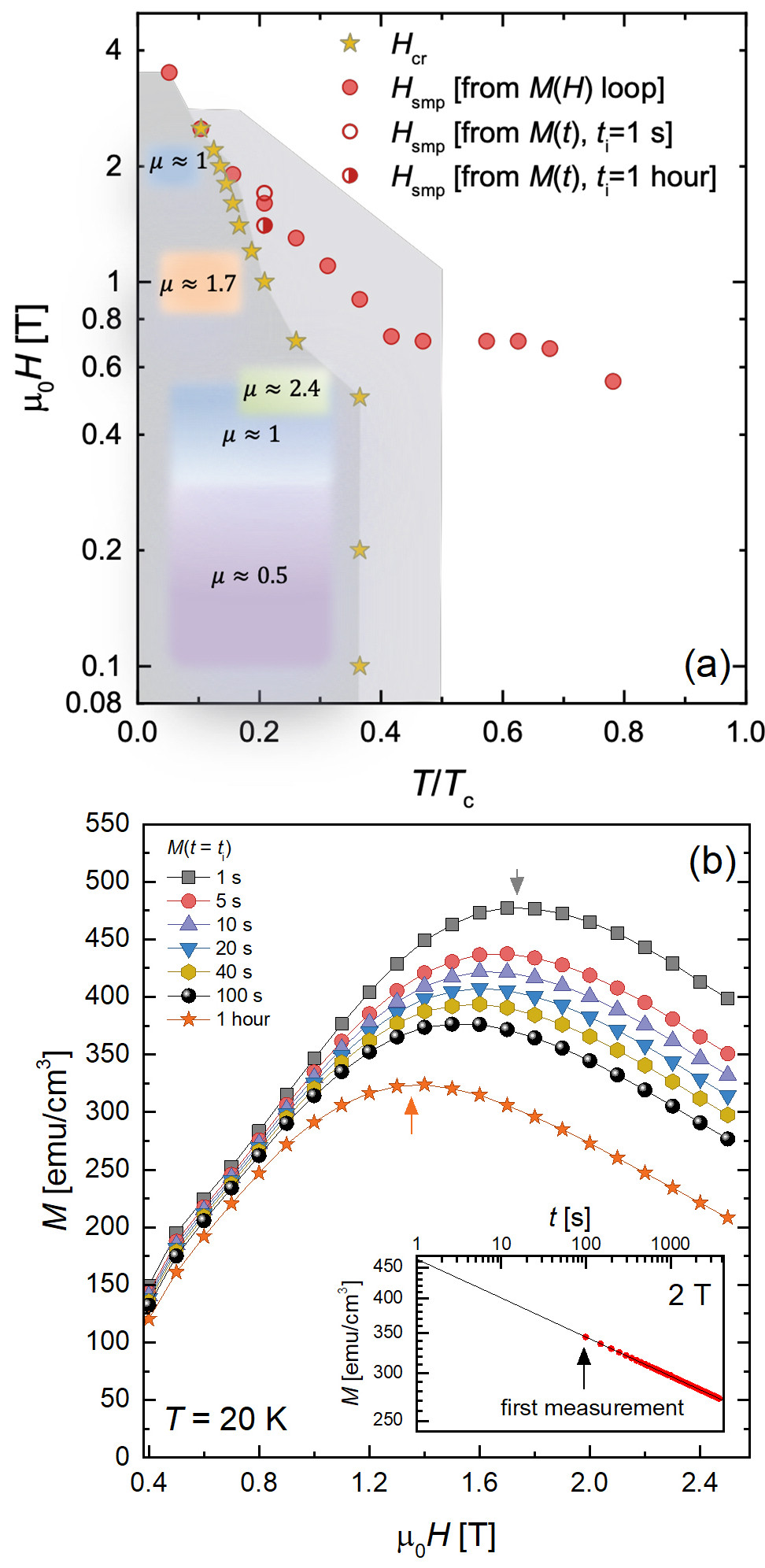}
\caption{(a) Vortex phase diagram determined from creep measurements overlaid with position of second magnetization peak. (b) Upper branch of $M(H)$ loop at $T = 20\textnormal{ K}$ constructed from magnetic relaxation data, where $M(t_i)$ represents the magnetization a time $t_i$ after the critical state was formed.  The inset demonstrates the extraction technique and how the first measurement collected by the magnetometer occurs at approximately $t_i$=100 s.}\label{Figure8}
\end{figure}

To explore this issue, we overlay our measurements of $H_{smp}$ and $H_{cr}$ in the phase diagram in Fig.\ \ref{Figure8}(a). Figure \ref{Figure6}(d)
show examples of how $H_{cr}$ was extracted from $U^*(1/J)$. At low temperatures $T/T_c < 0.2$, the appearance of the SMP coincides with the elastic-to-plastic crossover whereas $H_{cr} < H_{smp}$ at higher temperatures. Given this discrepancy, we now consider the previous argument that fast creep rates make measurements of $H_{smp}$ from loops inaccurate.

As described in Sec. IIIB, we typically create a magnetization loop by measuring $M$ once at each field as the field is ramped up in steps.  Constructing the loop instead from magnetic relaxation data enables us to set a consistent time scale for all $M$ values.  We achieve this by extracting $M$ from a linear fit to $\log M - \log t$ at a predetermined time $t_i$ after formation of the critical state, exemplified in the inset to Fig.\ \ref{Figure8}(b).  This also allows us to estimate $M$ before the first measurement.  The main panel shows how $M(H)$ changes with $t_i$. We find that a faster measurement increases $H_{smp}$, moving it away from $H_{cr}$, contrary to the observation in Pissas \textit{et al.} \cite{Pissas1999}. 

Conversely, the time scale associated with our determination of the elastic-to-plastic crossover is arguably the 1 hour duration of our creep measurements, therefore, an appropriate comparison requires $H_{smp}$ to be determined from $t_i = 1 \textnormal{ hour}$. As shown in Figs.\ \ref{Figure8} and \ref{Figure8}(b), though this reduces $H_{smp}$, it remains significantly larger than $H_{cr}$.  We can therefore conclude that the elastic-to-plastic crossover is not the source of the SMP at high temperatures.

\section{IV. Conclusions}

To summarize, we have studied the field- and temperature- dependent magnetization and vortex creep in an HgBa$_2$CuO$_{4+\delta}$ single crystal to understand the effects of anisotropy on vortex dynamics in superconductors. We reveal glassy behavior involving collective creep of bundles of 2D pancake vortices over a broad range of temperatures and fields as well as temperature- and time-tuned crossovers from elastic dynamics to plastic flow.  The isothermal magnetization loops exhibit distinct second magnetization peaks that have also been observed in previous studies of Hg1201, and $H_{smp}(T)$ decays exponentially at low temperatures then exhibits an abrupt change in behavior above $T/T_c = 0.5$.  The origin of the second magnetization peak in superconductors can be controversial, and is often attributed to an elastic-to-plastic crossover. Here we clearly show that the second magnetization in Hg1201 is not caused by an elastic-to-plastic crossover at $T/T_c >0.2$ and occurs within the plastic flow regime.

\section{Acknowledgements}
\begin{acknowledgments}
We would like to thank V. Vinokur for useful discussions regarding the plastic flow regime. This material is based upon work supported by the National Science Foundation under Grant No. 1905909 (data analysis and manuscript composition) and the U.S. Department of Energy, projects “Towards a Universal Description of Vortex Matter in Superconductors” (experimental measurements), "Science at 100 T" (crystal synthesis) and "Quantum Fluctuations in Narrow Band Systems" (crystal synthesis). 
\end{acknowledgments}

\bibliography{Hg1201bib}

\end{document}